\documentclass[iop]{emulateapj_pdftex}
\usepackage{amsmath,dsnr,epsfig,ifthen,natbib}
\newcommand{\hst}{{\it HST}}
\newcommand{\vfifty}{\ifmmode v_{50\%}\else $v_{50\%}$\fi}
\newcommand{\vtsig}{\ifmmode v_{98\%}\else $v_{98\%}$\fi}

\defcitealias{rupke13a}{RV13}

\shorttitle{A Resolved Molecular Outflow in a Buried QSO}
\shortauthors{Rupke \& Veilleux}

\begin{document}

\journalinfo{The Astrophysical Journal Letters, in press}
\slugcomment{Received 2013 July 3; accepted 2013 August 21}

\title{Breaking the Obscuring Screen: A Resolved Molecular Outflow in
  a Buried QSO}

\author{David S. N. Rupke} \affil{Department of Physics, Rhodes
  College, Memphis, TN 38112} \email{drupke@gmail.com}

\author{and Sylvain Veilleux} \affil{Department of Astronomy and Joint
  Space-Science Institute, University of Maryland, College Park, MD
  20742}

\begin{abstract}
  We present Keck laser guide star adaptive optics observations of the
  nearby buried QSO F08572$+$3915:NW. We use near-infrared integral
  field data taken with OSIRIS to reveal a compact disk and molecular
  outflow using \paa\ and \hmol\ rotational-vibrational transitions at
  a spatial resolution of 100~pc. The outflow emerges perpendicular to
  the disk into a bicone of one-sided opening angle 100$^\circ$ up to
  distances of 400~pc from the nucleus. The integrated outflow
  velocities, which reach at least $-$1300~\kms, correspond exactly to
  those observed in (unresolved) OH absorption, but are smaller
  (larger) than those observed on larger scales in the ionized
  (neutral atomic) outflow. These data represent a factor of $>$10
  improvement in the spatial resolution of molecular outflows from
  mergers/QSOs, and plausibly represent the early stages of the
  excavation of the dust screen from a buried QSO.
\end{abstract}


%%%%%%%%%%%%%%%%%%%%%%%%

\section{INTRODUCTION} \label{sec:intro}

Wide angle, large scale outflows in quasi-stellar objects (QSOs) that
are driven by black hole accretion energy are a key element of models
of major galaxy mergers. In long-standing merger models, such outflows
act as feedback on star formation and AGN. Furthermore, they may
transform QSOs buried in dusty molecular disks into true QSOs by
``breaking the obscuring screen'' of dust
\citep[e.g.,][]{sanders88a,hopkins05a}.

Large scale, AGN-driven outflows in major mergers have recently been
discovered in the ionized, neutral, and molecular gas phases. High
velocity outflows in nearby ultraluminous infrared galaxies (ULIRGs)
show evidence for acceleration by an AGN in those galaxies where one
is present (\citealt{rupke11a,westmoquette12a,rupke13a}, hereafter
RV13; \citealt{veilleux13a}). The mechanical luminosities of these
outflows (10$^{43}$ erg s$^{-1}$, and $0.002-004~L_{AGN}$) are
inconsistent with driving by a starburst alone but consistent with
recent models of AGN feedback (\citealt{hopkins10a,rupke11a};
RV13). The peak velocities in systems that contain an AGN are in the
range $1500-3500$~\kms, while the velocities in galaxies without an
AGN peak near 1000~\kms\ \citepalias{rupke13a}.

Outflows in mergers and QSOs also have a molecular gas phase, as
observed in far infrared (FIR) {\it Herschel} spectra
\citep{fischer10a,sturm11a,veilleux13a}. OH absorption lines show
velocities over 1000~\kms\ in some galaxies containing an AGN. In
several major mergers and/or QSOs, extended, high-velocity CO emission
is also observed
\citep{feruglio10a,aalto12a,cicone12a,cicone13a}. This gas has
velocities up to $800-1000$~\kms, and the sizes of these outflows are
$\ga$1~kpc. However, these data were obtained with beam sizes similar
to or larger than the inferred spatial extent of the CO emission.

In this paper, we present integral field spectroscopy (IFS), aided by
laser guide star adaptive optics, of the major merger
F08572$+$3915. This interacting system shows high velocity, large
scale outflows in neutral, ionized, and molecular gas
(\citealt{sturm11a}; RV13; \citealt{cicone13a}). The FIR luminosity is
concentrated in the northwest (NW) nucleus of the system and is
powered predominantly by a QSO \citep{soifer00a,armus07a,veilleux09a},
which is heavily obscured at all wavelengths (see detailed discussion
and references in Section 4.1.1 of RV13).

The spatial resolution of these observations is a factor of 7 higher
than in seeing-limited optical observations (0\farcs6 seeing; RV13),
and a factor $\sim$30 higher than in millimeter observations
(3\farcs1$\times$2\farcs7 beam; \citealt{cicone13a}). We can thereby
probe the spatial structure of the outflow, using near-infrared (NIR)
recombination lines and \hmol\ rotational-vibrational lines, at 100~pc
scales. NIR \hmol\ lines are an excellent tracer of the molecular
phase in the M82 wind \citep{veilleux09c}. 100~pc approaches the
scales at which AGN energy may couple to the outflow (e.g., similar to
the radial location of some UV and X-ray absorbers;
\citealt{crenshaw12a}).

In Section \ref{sec:obs}, we discuss the observations and data
analysis. We present the results in Section \ref{sec:res}, and discuss
them in Section\,\ref{sec:dis}.

\begin{figure*}[t]
  % \plotone{f1.eps}
  \centering \includegraphics[width=6.5in]{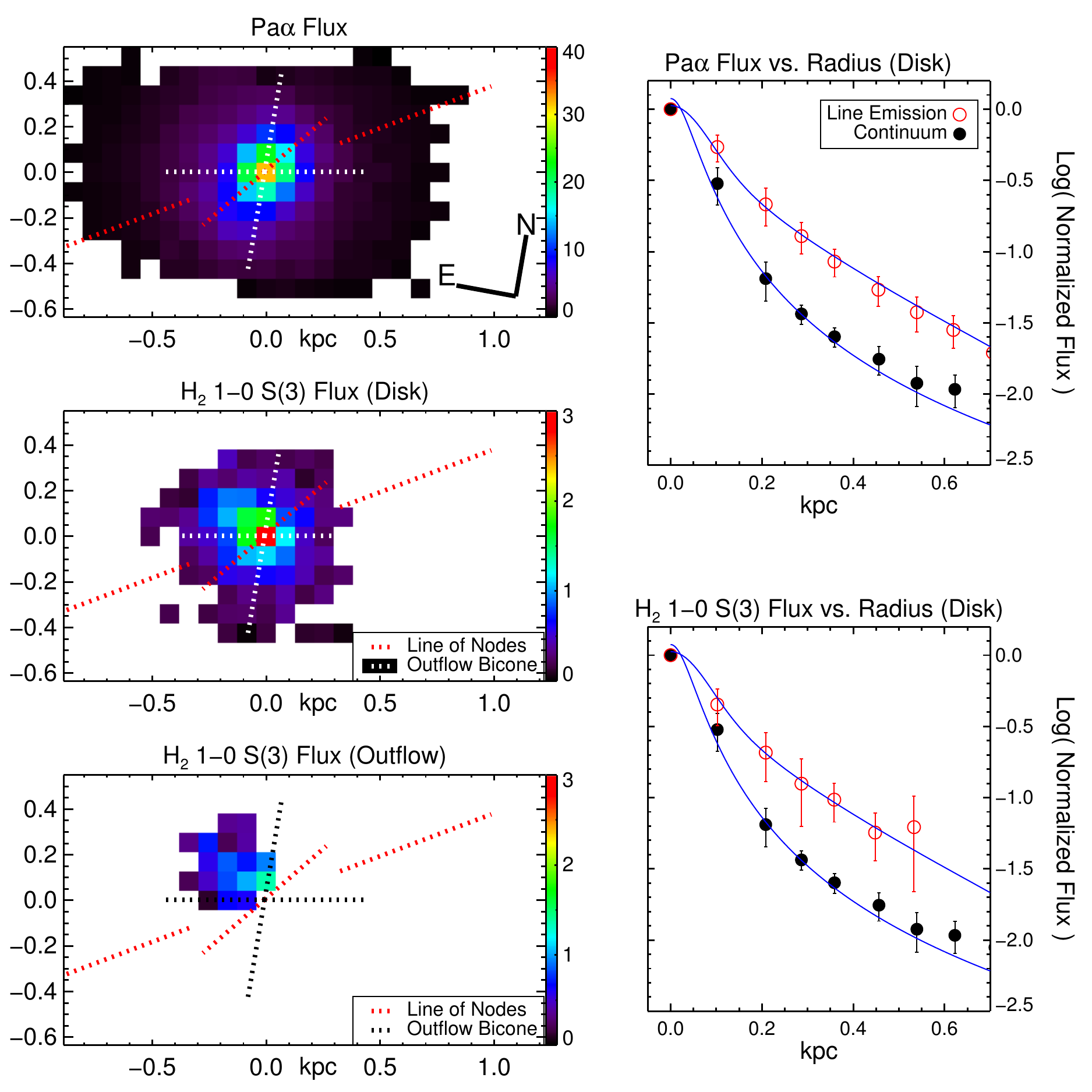}
  \caption{\small ({\it Left}) Flux maps of the gas disk and outflow
    in F08572$+$3915:NW, in units of
    10$^{-16}$~erg~s$^{-1}$~cm$^{-2}$. The top two maps shows \paa\
    and \hozsth\ in the disk and the bottom map shows \hozsth\ in the
    outflow. The red dotted lines show the approximate lines of nodes
    of the inner (140$^\circ$ east of north) and outer (120$^\circ$)
    disks, while the black and white dotted lines outline a possible
    biconical shape for the wind (of one-sided opening angle
    100$^\circ$). (For comparison, continuum and dust images appear in
    Figure~\ref{fig:dust}.) ({\it Right}) Flux vs. radius in 0\farcs07
    radial bins. Black filled circles show the $2.0-2.2\micron$
    continuum; red open circles show line emission; and error bars
    represent dispersion. The blue line through the continuum data is
    a Moffat fit (FWHM $=$ 0\farcs09). The blue line through the
    emission line data is a Gaussian $+$ exponential model.}
  \label{fig:flux}
\end{figure*}

\section{OBSERVATIONS AND DATA ANALYSIS} \label{sec:obs}

We observed F08572$+$3915:NW using the OH-Suppressing Infra-Red
Imaging Spectrograph (OSIRIS; \citealt{larkin06a}) on Keck~I on 31
January 2013 UT. We used the 0\farcs035 lenslet array and the Kbb
filter to achieve spectral coverage of 1.97 to 2.38$\mu$m. We took
eight 900~s on-source exposures interleaved with two 900~s sky frames
and dithered to achieve a mosaiced field of view (FOV) of
1\farcs0$\times$2\farcs9. The A0V star HD 95126 served as a telluric
standard.

We used v3.2 of the OSIRIS pipeline to sky subtract each on-source
exposure and produce a reduced data cube. We then manually aligned the
exposures by fitting the galaxy nucleus. We used the pipeline to
mosaic the data, combining the exposures with the MEANCLIP algorithm
and a 2.35$\sigma$ threshold for rejection. We removed recombination
lines from the telluric spectrum and normalized it with a 9480~K
blackbody before dividing the science data by the telluric
spectrum. We binned the data cube into 0\farcs07 square spaxels. To
flux calibrate the data, we compared Vega flux densities
\citep{tokunaga05a} to the measurements of our telluric standard, with
the normalization between the two stars determined from 2MASS fluxes.

\begin{figure*}[top]
  % \plotone{f2.eps}
  \centering \includegraphics[width=6.5in]{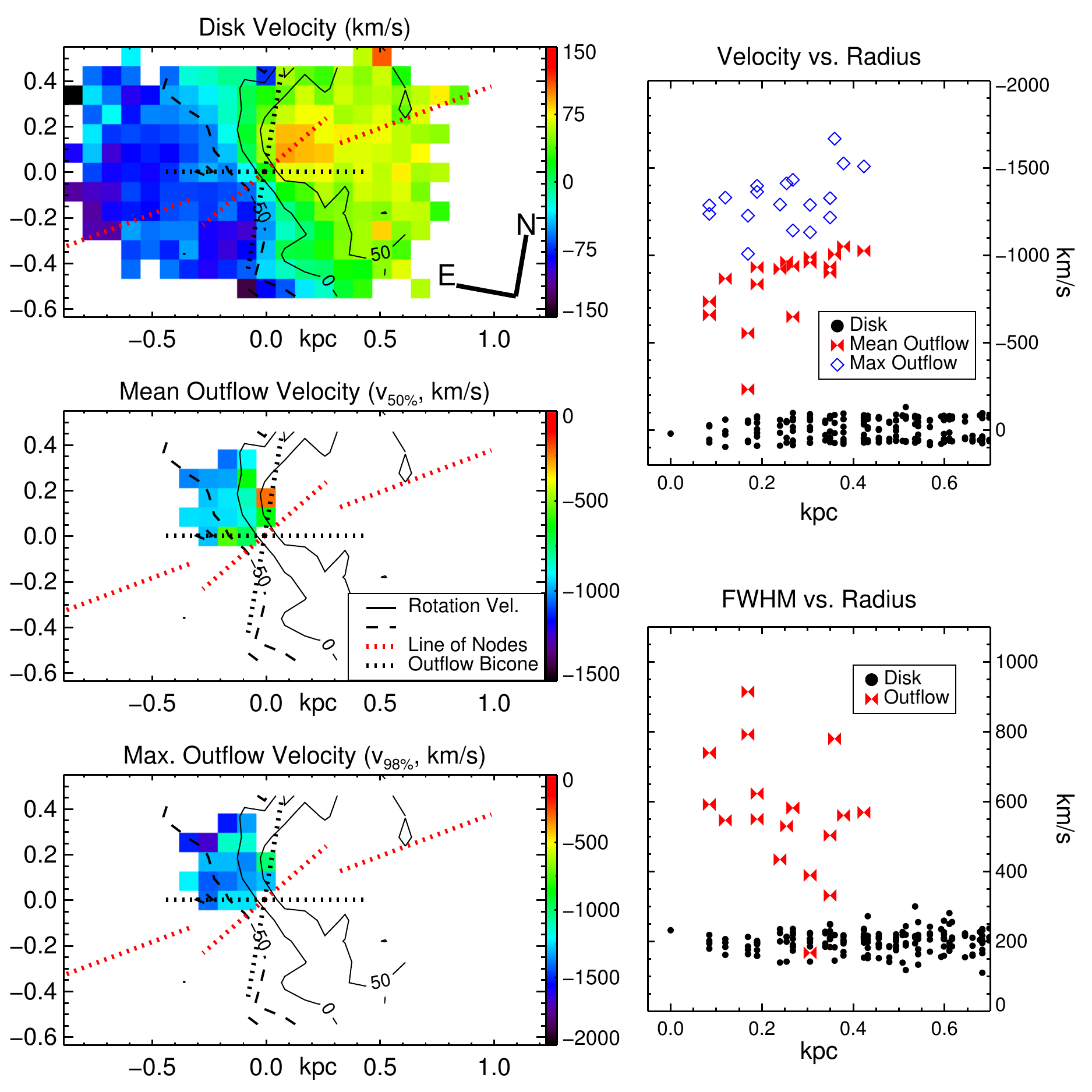}
  \caption{\small ({\it Left}) Velocity maps of rotation and outflow
    in F08572$+$3915:NW, with respect to $z=0.0583$. From top to
    bottom: ionized gas rotation, mean outflow velocity (\vfifty) in
    \hmol, and maximum outflow velocity (\vtsig) in \hmol. The red and
    black dotted lines are the same as in Figure~\ref{fig:flux}. The
    blueshifted component extends along the minor axis of the nuclear
    disk. ({\it Right}) Velocity and FWHM vs. radius. Black filled
    circles represent velocity and FWHM of rotating components (\paa),
    red filled hourglasses show mean outflow velocity and outflow FWHM
    (\hmol), and blue open diamonds show maximum outflow velocity. The
    outflow is more blueshifted with increasing radius, while the FWHM
    declines.}
  \label{fig:vel}
\end{figure*}

We fitted the data using UHSPECFIT \citep{rupke10b}. In each spaxel we
simultaneously fitted a third order polynomial and a set of Gaussians
to the continuum and emission lines, respectively. The emission lines
we fitted were \paa, \brd, \brg, \ion{He}{1} 1.87$\mu$m, \ion{He}{1}
2.06$\mu$m, \hozsth, \hozstw, and \hozso. We used one Gaussian for the
recombination lines and one or two for the \hmol\ lines. The
recombination lines were constrained to have the same redshift and
linewidth within each spaxel. The same was done for the molecular
lines, though this redshift and linewidth could be different from
those of the recombination lines. \htwo\ components were required to
be detected in \paa\ at the 3$\sigma$ level, while \hmol\ components
were required to be detected in at least two separate lines at
3$\sigma$ or an individual line at 5$\sigma$. The second velocity
component in the molecular lines, if present in a given spaxel at a
significant level, was found to be always blueshifted, broader, and
weaker in peak flux than the first component. For this broad
component, the fit was constrained only by the S(3) and S(1) lines, as
the S(2) line was not detected in integrated spectra
(\S\,\ref{sec:obs}). Furthermore, this second component was allowed in
a few spaxels on the basis of its clear presence after visual
inspection, even if it was slightly under the quantitative
significance criterion.

\begin{figure*}[t]
  % \plotone{f3.eps}
  \centering \includegraphics[width=6.5in]{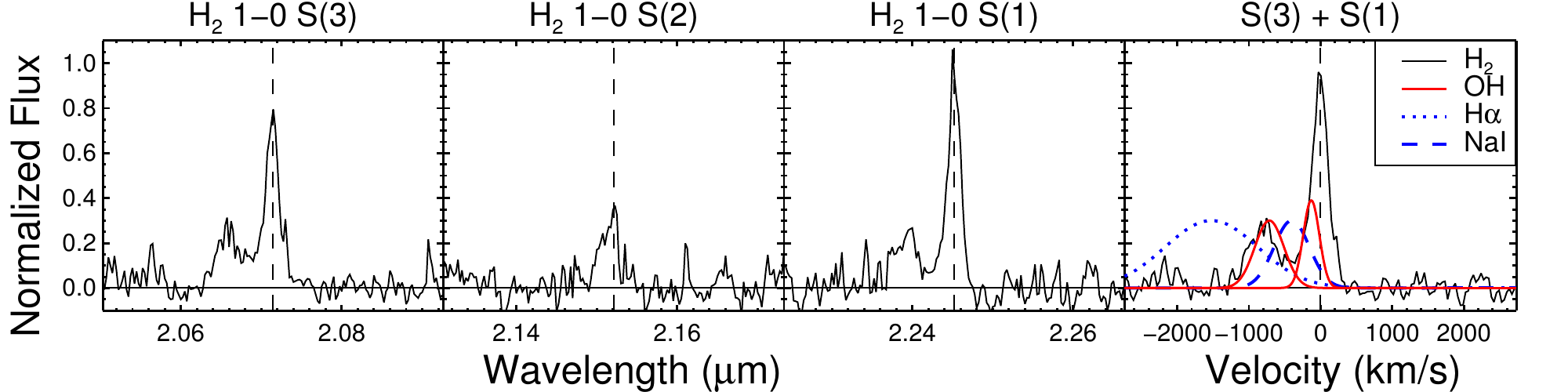}
  \caption{NIR \hmol\ profiles in F08572$+$3915:NW. From left to
    right: \hozsth, \hozstw, \hozso, and the sum of the \hozsth\ and
    \hozso\ emission line profiles, summed over those spaxels in which
    a broad, blueshifted component is detected. In the rightmost
    panel, we overplot the two components of the OH velocity profile
    as solid red lines \citep{veilleux13a} and the spatially-averaged
    \ha\ and \nad\ outflow velocity profiles as dotted and dashed blue
    lines \citepalias{rupke13a}. There is excellent correspondence
    between the high-velocity \hmol\ and OH components. The ionized
    gas is overall more blueshifted. The neutral gas is lower in
    velocity, but it arises from a different location.}
  \label{fig:spec_of}
\end{figure*}

A radial Moffat fit to the $2.0-2.2\micron$ continuum data yielded a
full width at half maximum (FWHM) of 0\farcs09, corresponding to
100~pc (Figure \ref{fig:flux}). The fit accounts for sampling effects
in the central spaxel, and is consistent with the continuum being an
unresolved point source. \citet{scoville00a} also find that the
nucleus of this system is unresolved in the NIR with the {\it Hubble
  Space Telescope} (\hst).

\section{RESULTS} \label{sec:res}

The \htwo\ and \hmol\ lines in F08572$+$3915:NW reveal the presence of
a compact gas disk. Figure~\ref{fig:flux} shows flux maps in \paa\ and
\hozsth\ and binned radial profiles. The data are consistent with a
compact source plus an extended component. We model this superposition
in radial space as a Gaussian (FWHM $\sim$ 100~pc) plus an exponential
(scale length $\sim$ 200~pc), after convolving each with the continuum
point spread function. In our model, the Gaussian and exponential have
a peak flux ratio of 3:1. This model, though azimuthally-averaged,
illustrates that the central gas concentration is compact. If
$100-200$~pc is a characteristic radius for the nuclear gas disk, then
it is more compact than many other ULIRG disks
\citep{downes98a}. However, systematic effects may impact the
determination of the disk size from NIR tracers (e.g., the heavy
nuclear extinction in this galaxy). CO or other cold molecular gas
observations would be more conclusive.

The disk kinematics are shown in Figure~\ref{fig:vel}. \htwo\ and
\hmol\ tracers of the disk yield similar results, but the \htwo\
tracers have higher signal-to-noise ratio and allow the disk to be
traced to larger radius. We thus show only the \paa\ kinematics. A
rotating ionized disk of at least 3 kpc radius had already been
detected \citepalias{rupke13a}. The systemic redshifts determined from
\paa\ ($z = 0.0583$) and optical emission lines
\citepalias[0.0584;][]{rupke13a} are identical within the errors.

The kinematics at small scales show a slight misalignment of the line
of nodes compared to that at larger scales. The \ha\ velocities at
radii $>$1~kpc have an apparent disk line of nodes of
$(120\pm10)^\circ$ \citepalias{rupke13a}. At smaller scales the
isovelocity contours twist. The compact \paa\ disk has a line of nodes
of $(140\pm10)^\circ$ at radii $\la$200~pc. This kinematic
misalignment could arise from a disk warp or under the influence of
stellar structures such as a bar or oval distortion
\citep{emsellem06a,riffel11a}.

The second molecular gas component is broad and blueshifted
(Figures~\ref{fig:vel} and \ref{fig:spec_of}). It extends directly
along the galaxy minor axis from 100 to 400~pc NE of the nucleus. The
mean velocity in this component (which we label \vfifty, meaning that
50\%\ of the gas covered by a spaxel is less blueshifted than this
velocity) decreases with increasing radius from $-$700~\kms\ at radii
$<$200~pc to $-$1000~\kms\ at radii $>$300~pc, while the FWHM declines
from 700~\kms\ to 500~\kms. To describe the most blueshifted
velocities, we use $\vtsig \equiv \vfifty - 2\sigma$, meaning that
98\% of the gas covered by a spaxel is less blueshifted than this
velocity. The value of \vtsig\ remains fairly constant with radius,
with a mean of $-$1300~\kms\ and a range of $-$1000 to
$-$1700~\kms. However, given the noise in the line wings in individual
spaxels, the actual velocities may not reach $-1700$~\kms. The
integrated line profile reaches velocities of $-1300$~\kms\
(Figure~\ref{fig:spec_of}). We conclude that the most blueshifted warm
\hmol\ gas velocity is between $-1300$ and $-1700$~\kms.

\begin{figure*}[t]
  % \plotone{f4.eps}
  \centering \includegraphics[width=6.5in]{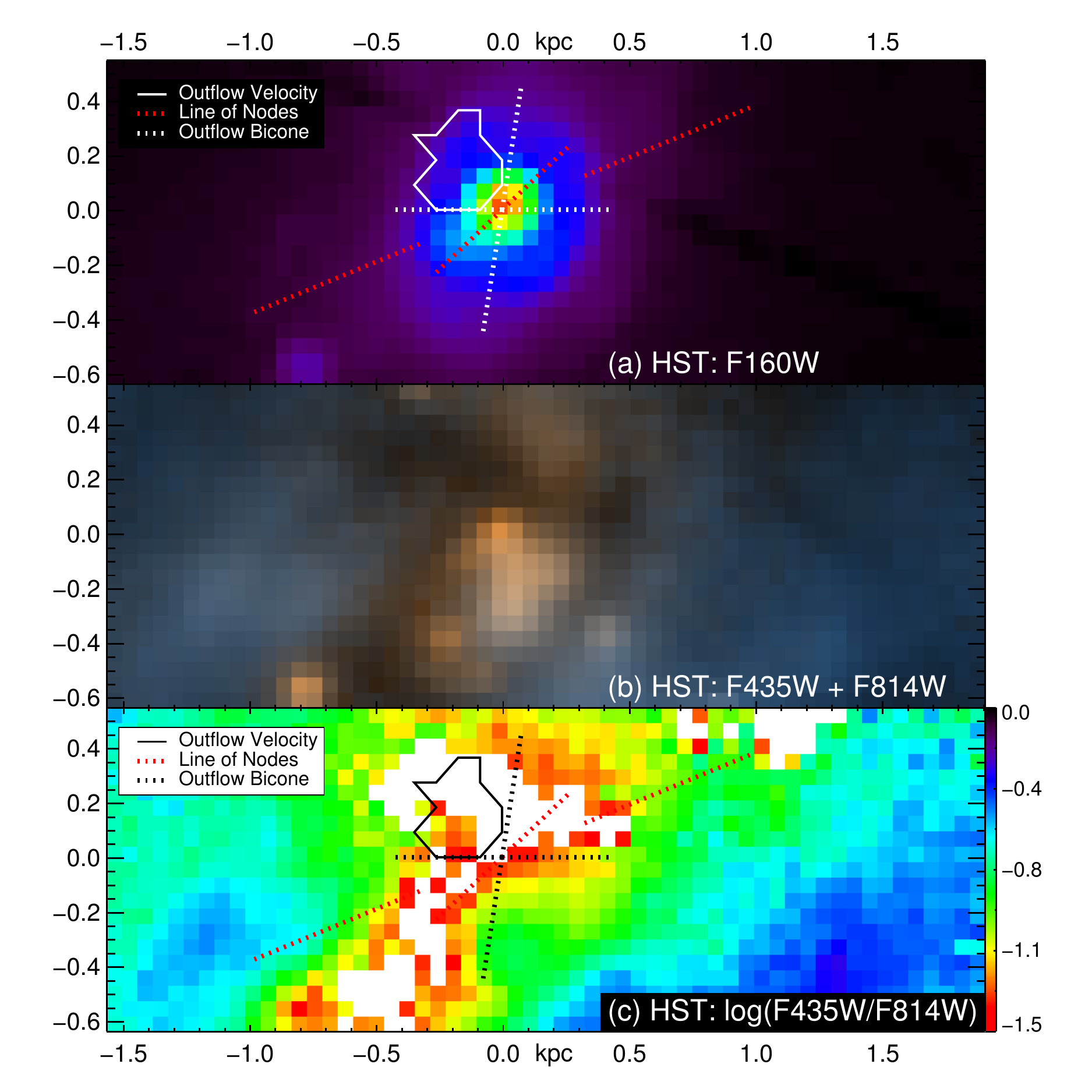}
  \caption{{\it Hubble Space Telescope} images of
    F08572$+$3915:NW. From top to bottom, the images are NICMOS F160W,
    ACS WFC F435W$+$F814W, and log(F435W/F814W). In the bottom panel,
    white represents areas where the extinction is so heavy that the
    galaxy is not detected in F435W. In the top and bottom panels, we
    overlay the contours of the nuclear outflow velocity field, the
    disk line of nodes, and an estimate of the outflow's opening
    angle. The nuclear outflow is coincident with a plume of dust
    absorption.}
  \label{fig:dust}
\end{figure*}

F08572$+$3915:NW was previously known to host a molecular outflow,
based on blueshifted OH absorption lines
\citep{sturm11a,veilleux13a}. In Figure~\ref{fig:spec_of}, we plot the
two components of the OH profile (as fit in \citealt{veilleux13a}) on
top of the \hmol\ profile (integrated over the outflow region). There
is excellent agreement in the velocities of the blueshifted OH and
\hmol\ components. The low-velocity OH component traces the blue wing
of the narrow \hmol\ component.

We conclude that the broad, blueshifted \hmol\ component in this
galaxy is a minor axis molecular outflow. We observe only one side of
the outflow, and infer that the counter-propagating side of the flow
is hidden by the galaxy disk. The \hmol\ line intensity declines away
from the galaxy nucleus and is highest through the bisector of the
outflow cone (Figure~\ref{fig:flux}). The increase of average velocity
with radius (Figure~\ref{fig:vel}) may result from a velocity
segregation (high velocity gas reaching large radii more quickly than
low velocity gas), rather than reflecting acceleration of the flow
\citep{dallavecchia08a}.

The (one-sided) wind opening angle inferred from our data is
$(100\pm10)^\circ$ (Figure~\ref{fig:vel}). This is smaller than the
typical one-sided opening angle inferred from OH or \nad\ surveys
\citep{veilleux13a,rupke05b}. These surveys yield a detection rate of
70\%, which imply a (one-sided) conical opening angle of
145$^\circ$. Similarly, preliminary modeling of the OH line in
F08572$+$3915:NW yields an opening angle of $\sim$150$^\circ$
\citep{sturm11a}.

The \hozsth/\hozso\ and \hozsth/\hozstw\ line ratios in the disk yield
median temperatures of 1700~K and 1300~K, respectively. If this
emission is thermal in origin, then these temperatures could be made
consistent with a reduction in the ortho-to-para ratio from 3.0 to
$2.1-2.2$ \citep{smith97a}. The \hozsth/\hozso\ line ratio in the
outflow shows a median excitation temperature of 2370~K. This
temperature is higher than in the disk (a K-S test shows that the disk
and outflow temperature distributions differ at the 96\%\ level).

The next levels of rotational-vibrational excitation of \hmol\ in our
wavelength range are the 1$-$0 S(4) line and the 2$-$1 lines. Systemic
1$-$0 S(4) is detected in our spectra, but it lies at the bottom of a
strong telluric feature. In integrated spectra, it has a similar flux
to the 1$-$0 S(2) line. The brightest 2$-$1 line is likely H$_2$ 2$-$1
S(3) at 2.07~\micron. The integrated spectrum yields an upper limit of
0.2 for the \hmol~2$-$1 S(3)/\hozso\ line ratio.

\section{DISCUSSION} \label{sec:dis}

This data set on F08572$+$3915:NW is significant for two
reasons. First, it shows a high spatial resolution view of a molecular
outflow driven by a QSO in a galaxy merger. It is thus one of the
handful of such molecular outflows that is resolved, and the only one
resolved at sub-kpc scales. Second, it may be an example of AGN
feedback in action: a deeply buried QSO that is in the process of
removing the dusty molecular gas that obscures it.

Neutral atomic, ionized, and molecular gas outflows have been detected
in other studies of F08572$+$3915:NW (\citealt{sturm11a}; RV13;
\citealt{cicone13a}). At a resolution of 1~kpc, the ionized gas is
elongated along the minor axis of the galaxy, and it has the highest
outflow velocities (peaking at 3350~\kms) among six major mergers
studied with IFS \citepalias{rupke13a}. As observed in both single
aperture spectra of OH and in CO emission with a resolution of several
kpc, the outflowing gas has a peak velocity of $700-800$~\kms\ and a
maximum velocity (\vtsig) of $1100-1200$~\kms\
\citep{sturm11a,veilleux13a,cicone13a}. Finally, high-excitation CO
absorption has been detected at blueshifts up to $-400$~\kms\
\citep{geballe06a,shirahata13a}.

Besides the improved spatial resolution, this data set represents one
of the few resolved observations of a molecular outflow in major
mergers or QSOs. Herschel observations of molecular outflows in ULIRGs
are spatially unresolved \citep{sturm11a,veilleux13a}, and previous CO
observations of ULIRGs or QSOs have resolved molecular outflows in
only a few cases
\citep{feruglio10a,aalto12a,cicone12a,feruglio13a,cicone13a}. Most of
the CO outflows in these systems are asymmetric, with the blue and/or
red wing extended to one side, while in NGC 6240 the extended CO gas
is in filaments that are co-spatial with extended warm and hot ionized
gas structures.

In F08572$+$3915:NW, we determine that the molecular wind is
collimated by the nuclear disk along the minor axis of the system. The
same was found for the ionized gas in this galaxy, and more generally
in other mergers on scales up to 2~kpc \citepalias{rupke13a}. A
one-sided, limb-brightened superbubble is also seen in \hmol\ emission
in NGC~4945 \citep[e.g.,][]{moorwood96a,marconi00a}. However, rather
than being limb-brightened, the \hmol\ emission in the
F08572$+$3915:NW outflow is concentrated near the outflow axis
(Figure~\ref{fig:flux}).

Previous authors have concluded that the buried QSO in
F08572$+$3915:NW plays a significant role in powering the outflow
(\citealt{sturm11a}; RV13; \citealt{veilleux13a}). This conclusion was
based on the high velocities observed (greater than outflow velocities
observed in similar systems with no detected AGN), the high energy of
the outflow (which requires an unreasonably high thermalization
efficiency if the starburst alone powers the outflow), and the high
momentum of the outflow.

The present data strengthen the case for a significant contribution
from an AGN. Given its size, the outflow cannot emerge from a region
much larger than $\sim$200~pc in diameter. It is consistent with
emerging from a region that is even smaller. The mid-infrared
continuum source in this system is very small (size $<$100~pc;
\citealt{imanishi11a}), and it is dominated by the buried AGN
\citep{veilleux09a}. Furthermore, the inner gas disk, as traced by
\paa\ and NIR \hmol\ lines, is apparently concentrated within the
central $100-200$~pc in radius (\S~\ref{sec:res}).

Given that we only detect the outflow securely in two molecular lines
(Figure \ref{fig:spec_of}), it is difficult to constrain the gas
excitation. However, the wind is more highly excited than the
disk. Typically, NIR and MIR \hmol\ emission in ULIRG disks is
consistent with originating in PDRs \citep{davies03a,higdon06a},
though shock excitation may also play a role
\citep{vanderwerf93a,zakamska10a}.

Strikingly, the warm and cold molecular phases share very similar
kinematics (Figure~\ref{fig:spec_of}). Thus, it is plausible that the
warm and cold molecular gas are cospatial. The warm \hmol\ could
reside in the outer regions of dense clouds that are being externally
heated.

To estimate the mass of warm \hmol\ in the outflow, we assume thermal
equilibrium at $T = 2370$~K (\S~\ref{sec:res}). The \hozsth\ flux is
$1.1\times10^{-15}$ erg~s$^{-1}$~cm$^{-2}$. The equations from
\citet{roussel07a} and the partition function from \citet{herbst96a}
give $N(\hmol) = 1.8\times10^{19}$ cm$^{-2}$ and
$M(\hmol,~\mathrm{warm}) = 5.2\times10^4$ \msun. This is a factor
$1.6\times10^3$ lower than the mass in the neutral atomic and ionized
phases of the wind \citepalias{rupke13a}. Based on our measured
$\vfifty = -1000$~\kms\ at 400~pc, the dynamical time is 0.4~Myr,
yielding a mass outflow rate of $dM/dt(\hmol,~\mathrm{warm}) =
0.13$~\smpy.

As determined from OH transitions, $dM/dt(\hmol)$ is
1000$^{+2900}_{-730}$~\smpy\ \citep{sturm11a}, while the ionized and
neutral gas outflow rate is 30~\smpy\ \citepalias{rupke13a}. To bring
the warm molecular outflow rates in line with other measured rates,
the ratio of total to warm molecular gas in the F08572$+$3915:NW
outflow must be lower than in the M82 wind by factors of $10-100$
\citep{veilleux09c}. The different radiation environments and outflow
speeds in the two galaxies may cause this difference.

Dust is present in the outflow in the form of an optically-thick
filament that wraps around the edge of the \hmol\ emission (Figure
\ref{fig:dust}). This plausibly implies that the molecular outflow is
beginning to excavate dust from around the AGN, and will uncover the
QSO on short timescales. This molecular flow may thus be a missing
link between the buried and naked QSO phases in the classic major
merger timeline \citep{sanders88a,hopkins05a}. The outflow in the
nearby QSO Mrk~231, which has reached scales of several kpc
\citep{rupke11a} may be a prime example of the next phase: a true QSO
with a large scale galactic wind that has already done its work.

\acknowledgments The authors thank the referee for a helpful
report. D.S.N.R. was supported by a NASA Keck PI Data Award,
administered by the NASA Exoplanet Science Institute, and by a
Cottrell College Science Award. S.V. acknowledges NSF support under
grant AST-10009583. Data presented herein were obtained at the
W.M. Keck Observatory from telescope time allocated to NASA. The
Observatory was made possible by the generous financial support of the
W.M. Keck Foundation. The \hst\ observations described herein were
obtained from the Hubble Legacy Archive.

\bibliography{apj-jour,dsr-refs}

\end{document}